\documentclass[aps,superscriptaddress,preprint]{revtex4}
\usepackage{natbib}
\usepackage{graphicx}
\usepackage{amssymb}
\usepackage{multirow}
\usepackage{url}

\begin{document}

\title[Simulation benchmarks]{Simulation benchmarks for low-pressure plasmas: capacitive discharges}
\author{M. M. Turner}
\email{miles.turner@dcu.ie}
\affiliation{School of Physical Sciences and National Centre for Plasma
Science and Technology, Dublin City University, Dublin 9, Ireland}

\author{A. Derzsi}
\affiliation{Hungarian Academy of Sciences, Institute for Solid State
Physics and Optics, Wigner Research Centre for Physics, 1121 Budapest, Konkoly-Thege Miklos Str. 29-33, Hungary}

\author{Z. Donk\'o}
\affiliation{Hungarian Academy of Sciences, Institute for Solid State
Physics and Optics, Wigner Research Centre for Physics, 1121 Budapest, Konkoly-Thege Miklos Str. 29-33, Hungary}

\author{D. Eremin}
\affiliation{Lehrstuhl f\"ur Theoretische Elektrotechnik,
Fakult\"at f\"ur Elektrotechnik und Informationstechnik,
Ruhr-Universit\"at Bochum,
Universit\"atsstra\ss e 150,
44801 Bochum, Germany}

\author{S. J. Kelly}
\affiliation{School of Physical Sciences and National Centre for Plasma
Science and Technology, Dublin City University, Dublin 9, Ireland}

\author{T. Lafleur}
\affiliation{Laboratoire de Physique des Plasmas, Ecole Polytechnique, 
91128 Palaiseau, France}

\author{T. Mussenbrock}
\affiliation{Lehrstuhl f\"ur Theoretische Elektrotechnik,
Fakult\"at f\"ur Elektrotechnik und Informationstechnik,
Ruhr-Universit\"at Bochum,
Universit\"atsstra\ss e 150,
44801 Bochum, Germany}

\begin{abstract}
Benchmarking is generally accepted as an important element in
demonstrating the correctness of computer simulations.  In the modern
sense, a benchmark is a computer simulation result that has evidence
of correctness, is accompanied by estimates of relevant errors, and
which can thus be used as a basis for judging the accuracy and
efficiency of other codes.  In this paper, we present four benchmark
cases related to capacitively coupled discharges.  These benchmarks
prescribe all relevant physical and numerical parameters.  We have
simulated the benchmark conditions using five independently developed
particle-in-cell codes.  We show that the results of these simulations
are statistically indistinguishable, within bounds of uncertainty that
we define.  We therefore claim that the results of these simulations
represent strong benchmarks, that can be used as a basis for
evaluating the accuracy of other codes.  These other codes could
include other approaches than particle-in-cell simulations, where
benchmarking could examine not just implementation accuracy and
efficiency, but also the fidelity of different physical models, such
as moment or hybrid models.  We discuss an example of this kind in
an appendix.  Of course, the methodology that we have
developed can also be readily extended to a suite of benchmarks with
coverage of a wider range of physical and chemical phenomena.
\end{abstract}
\maketitle

\section{Introduction}

This paper takes a first step towards a suite of benchmarks that can
be used to evaluate computer simulations for low-temperature plasmas
operated in the low-pressure regime.  We say a first step because the
scope of the present benchmarks is limited to simulations of
capacitive discharges with a simple geometry and a simple chemistry.
This limitation, however, is not immediately important to our
purpose---which is to demonstrate a rigorous approach to benchmark
development that is readily generalized to more complex cases.
Although our initial focus is on particle-in-cell simulations, we aim
to supply a benchmark that is relevant to all practitioners of
low-temperature plasma simulation.  We begin with particle-in-cell
simulations because that approach directly solves the Boltzmann
equation, which is usually accepted as the most fundamental physical
description of a low-temperature plasma.  Other methods are either
equivalent in physical content to a particle-in-cell simulation, or
involve approximations that are not required in a particle-in-cell
simulation. Consequently, we maintain that the results of
particle-in-cell simulations are at least as accurate as those
produced by any other technique, assuming, of course, that the
numerical parameters are well chosen.  Our approach has three steps.
The first is to define physical conditions for the benchmarks that are
fully prescriptive.  The second is to show that, when applied to these
benchmark cases, several independently developed particle-in-cell
simulations give statistically indistinguishable results.  This step
requires that we also specify all the numerical parameters required by
the particle-in-cell simulations.  We take success in this step as a
demonstration that the codes are accurately implemented.  The last
step is to investigate the influence of the numerical parameters on
the simulation results, so that an estimate can be supplied of the
residual uncertainty that remains in the benchmark simulation data.
Thus we have simulation results for the benchmark cases that we claim
to be correct in a strong sense, and which have well-defined bounds of
uncertainty.  These data will be available as an electronic supplement
to this paper, and may be used as a basis for evaluating the accuracy
and efficiency of computer simulations.

There is a broader context for this initiative.  In recent years,
evidence has been accumulating that the accuracy of computer
simulations is not what one might hope.  For example, outright errors
in scientific computer programs have been shown to be common
\citep{hatton_how_1994,hatton_t_1997}, even in professionally
maintained codes, and informal benchmarking exercises typically show a
wider range of results than is easily accounted for, even when the
algorithms implemented are nominally identical (see
\citet{oberkampf_verification_2002} for examples from a variety of
fields).  Responses to these disconcerting observations include calls
for more rigour in developing and testing computer simulation
programs, through clearly articulated and published ``verification and
validation'' activities
\citep{hatton_how_1994,hatton_t_1997,oberkampf_verification_2002,oberkampf_verification_2008,roy_comprehensive_2011}.
Some communities have gone so far as to require evidence that computer
simulations are correct as a routine precondition of publication
\citep{roache_editorial_1986,freitas_editorial_1993,kim_journal_1993}.
This body of work has thus far had little impact on the
low-temperature plasma physics community, whose literature shows
slight evidence of interest in demonstrating the formal correctness of
computer simulations, or quantifying the errors in simulation results.
Partial exceptions are \citet{lawler_self-consistent_1999} in which
careful benchmarks for positive columns were developed, and the swarm
physics community, which has historically been deeply engaged in
questions of modelling accuracy
\citep{petrovic_measurement_2009}---but even there, modern ideas about
systematic verification and validation have yet had little influence.
Of course, many authors of low-temperature plasma physics computer
simulations doubtless test their codes extensively, but such testing
is of limited value while unpublished.

An element in a systematic verification and validation campaign may be
a ``benchmark'' \citep{oberkampf_verification_2008}.  Other kinds of
tests typically exercise only parts of a code, or introduce artificial
elements for the purposes of testing, but a benchmark is a test of the
code on a problem of practical interest.  Usually,
and almost necessarily, a benchmark problem can be solved only by
computer simulation.  We note that a ``benchmark'' in this sense is
rather more than an informal comparison of codes: The aim is to
demonstrate a solution of the benchmark problem that can be widely
accepted as ``correct.''  A powerful way of increasing confidence in a
benchmark of this kind is to repeat the calculation with several
independent codes, and show that the results ``agree.''  In general,
one may find a definition of ``agreement'' difficult to discover, but
for inherently probabilistic methods, such as particle-in-cell
simulations, we can helpfully define ``agreement'' in terms of
``statistical indistinguishability,'' as we do below.

In the remainder of this paper, we will present the physical
parameters characterizing the benchmarks, describe the
particle-in-cell codes that have been used in the comparison together
with a methodology for demonstrating statistical indistinguishability,
show the results obtained for each of the benchmark cases, and discuss
the residual numerical uncertainty in these data.  In an appendix, we
present solutions of the benchmark problem using a moment model.
These data define a reasonable expectation for the ability of such
models to reproduce the benchmark results.  We have not, however,
subjected the moment model to any searching critical examination, and
we are not advancing strong claims concerning the correctness of these
data.  This is in contrast to the particle-in-cell data, which we
assert are correct solutions of the benchmark problems within the error
bounds that we supply.  We conclude the paper with a short discussion
followed by some final remarks.

\section{Benchmark description}

We have chosen the basic benchmark conditions to represent the
experiments of \citet{godyak_measurement_1992}.  This is with a view
to future validation against experiments, and also has the advantage
of similarity with the earlier benchmark of
\citet{surendra_radiofrequency_1995}.  We note that this pioneering
benchmark falls into the category alluded to above, where the
difference between codes is larger than perhaps expected, and the
origin of the differences remains mysterious.  In the present work,
four benchmark cases have been selected, with parameters that are shown in
table~\ref{tab_param}.  We assume a discharge between two plane and
parallel electrodes, where the electrodes are normal to
the $x$ axis.  This is, therefore, a one-dimensional Cartesian model.
The space between the electrodes is filled with helium gas at a
density that is fixed for each benchmark case, at a temperature of
300~K.  Both the gas density and temperature are unchanging in space
and time.  A sinusoidally varying voltage is applied between the
electrodes at a frequency of 13.56~MHz, with the phase convention that
the voltage is zero at time zero.  The voltage amplitudes are
different for each of the four benchmark cases, and have been chosen
to give approximately the same current density amplitude of
10~A~m$^{-2}$ at each of the four pressure points.  The pressure
points have been chosen to span approximately the range of values that
can reasonably be used---below the lower limit, no discharge can be
sustained, while above the upper limit, the collisionality is such that
the conventional particle-in-cell approach is inappropriate.
We assume that fluxes of charged particles reaching the electrodes are
completely absorbed, and no secondary particles (electrons, for example)
are emitted.

\begin{figure}
\includegraphics[width=0.9\columnwidth]{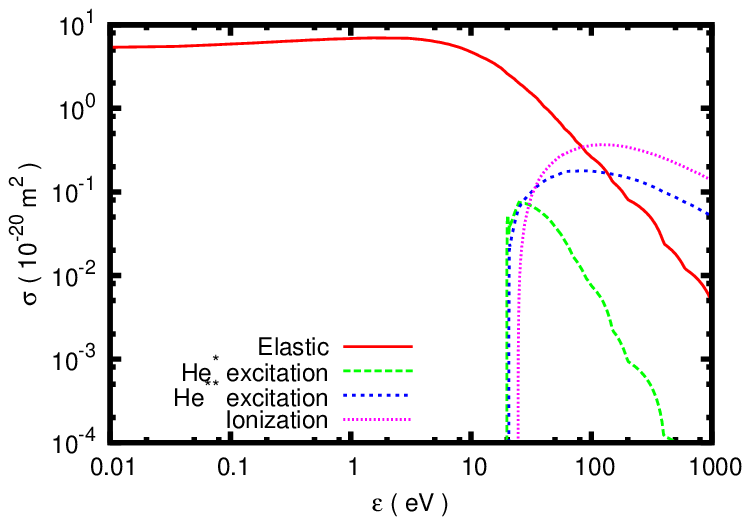}
\includegraphics[width=0.9\columnwidth]{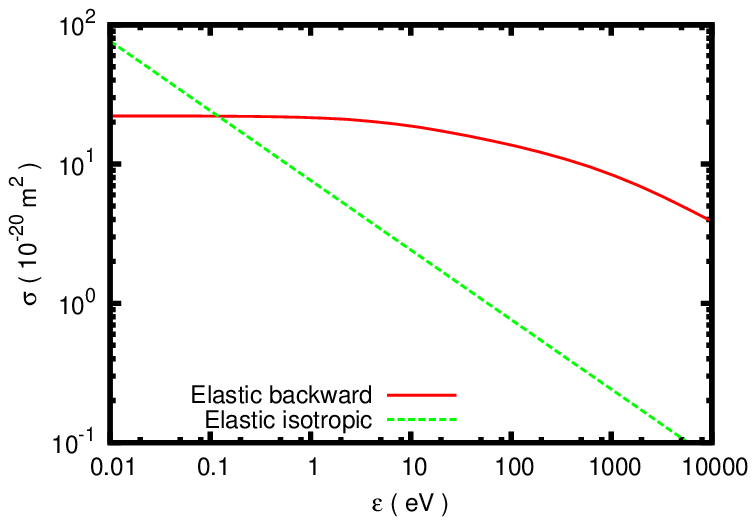}
\caption{Cross sections for electron-neutral collisions (upper panel)
and ion-neutral collisions (lower panel).  The energies shown in the
lower panel are expressed in the centre of mass frame.  \label{fig_xs}}  
\end{figure}

The benchmark assumes that the plasma is composed only of electrons
and helium monomer ions.  Collisional processes are therefore limited
to interactions between these species and neutral helium.  For
electron-neutral collisions, the cross section compilation known as
Biagi 7.1 is used \citep{biagi_cross_2004}.  This consists of an
elastic momentum transfer cross section, two excitation cross
sections, and an ionization cross section.  These cross sections
assume isotropic scattering in the centre of mass frame, and are
consistent with transport data when used in conjunction with such
scattering.  The benchmark therefore requires isotropic scattering for
all electron collision processes.  After an ionizing collision, the
residual energy is divided exactly equally between the primary and
secondary electrons.  For ion-neutral scattering, we adopt the
proposal of \citet{phelps_application_1994}, which approximates the
anisotropic scattering cross section as an isotropic scattering
component and a backward scattering component, both in the centre of
mass frame.  Such cross sections
for helium have been given as analytic expressions
\citep{phelps_compilation_2005}, which have been developed into tables
for use in the present benchmark.  These tables represent the cross sections
as functions of the centre of mass energy.  The tabulated cross sections for
both electrons and ions are to be interpolated linearly whenever
intermediate values are required.  In the (unlikely) event that data
are needed beyond the maximum tabulated energy, the last value in the
table is to be substituted.  Both sets of cross sections are available
as electronic supplements to this paper, and are shown for
reference in figure~\ref{fig_xs}. 
Further fundamental constants are to be represented by the 2006
CODATA values \citep{mohr_codata_2008}, expressed to at least four places of
decimals.

The procedure for running the benchmarks is the following.  The
simulation is initialized with the conditions specified in
table~\ref{tab_param}.  The system is then integrated in time for the
indicated interval, and an average is taken over a sub-interval at the
end of the calculation.  These time intervals are also specified in
table~\ref{tab_param}.  The average values obtained in this way are
the benchmark results.  In the discussion following, we focus on the
ion density distribution as the primary benchmark result.  This is because this
quantity has an unambiguous definition in any likely simulation procedure,
and is highly sensitive to both numerical effects and implementation
details.

\section{Simulation procedure}

Each of the simulation codes implements the classical particle-in-cell
algorithm with Monte Carlo collisions
\cite{hockney_computer_1981,birdsall_plasma_1991,birdsall_particle--cell_1991}.
As is well-known, in this approach the charged particles of the plasma
are represented by a set of so-called superparticles, which are much
smaller in number than the physical particles.  These superparticles
are immersed in self-consistently generated electric fields obtained
by solving Poisson's equation in a finite difference form on a uniform
spatial grid.  In the conventional expression of the method, the time
integration of particle trajectories uses the explicit leap-frog
scheme, which is second order accurate in the time step, $\Delta t$,
and the solution of the Poisson equation is second order accurate in
the cell size, $\Delta x$.  All the present codes use bi-linear
weighting for mapping grid quantities to particle positions and vice
versa
\citep{hockney_computer_1981,birdsall_plasma_1991,birdsall_particle--cell_1991}.
The particle-in-cell method is subject to stability and accuracy
conditions.  Generally accepted accuracy conditions for the cell size
and time step are
\begin{eqnarray} 
\omega_p \Delta t &\lesssim& 0.2 \label{eq_dt} \\
\frac{\lambda_D}{\Delta x} &\gtrsim& 2 \label{eq_dx},
\end{eqnarray} 
where $\omega_p$ is the plasma frequency and $\lambda_D$ is the Debye
length.  These results are derived from extensive computer experiments
\citep{hockney_computer_1981}.  Conditions \ref{eq_dt} and \ref{eq_dx}
can be combined to form a dependent constraint
\begin{equation}
\frac{\Delta t}{\Delta x}\sqrt{\frac{k_B T_e}{m_e}} \lesssim 0.4,
\end{equation}
showing that thermal electrons are displaced by much less than one
cell per time step when these conditions are satisfied.
There is a third numerical parameter, which controls the ratio between
the superparticle density and the physical particle density.  This is
often called the particle weight.  The particle weight is not defined
uniformly by the codes employed in this study.  The data given in
table~\ref{tab_param}, however, implicitly specify the particle weight
for all such definitions.  No universal rule exists for selecting the
particle weight, and indeed the evidence suggests that a wide
range of values may be appropriate to different contexts
\citep{turner_kinetic_2006}, so that a ``rule-of-thumb'' 
cannot be given at present.  The numerical parameters for particle-in-cell
simulations given in table~\ref{tab_param} have been chosen
conservatively according to the information available, and after some
experimentation.  This topic will be revisited below when we discuss
numerical uncertainty in the benchmark results.

Monte Carlo collisions are handled in the usual way
\citep{birdsall_particle--cell_1991,vahedi_monte_1995,donko_particle_2011},
by testing for collisions once per time step.  This procedure is
appropriate when the collision frequency is small compared with the
plasma frequency, which is the usual situation in low-pressure
discharges.  In general, the collision frequency for each particle
species varies in some arbitrary way with relative speed.  However,
appreciable algorithmic simplification is achieved by adopting the
so-called null collision method, in which a fictitious collision
process is introduced in order to render the total collision frequency
a constant for each particle species, denoted by $\nu_{e,i}$.  One
then has a constant collision probability per time step for each
species:
\begin{equation}
{\cal P}_{e,i} = 1 - \exp( -\nu_{e,i} \Delta t ) \approx \nu_{e,i} \Delta t.
\end{equation}
Once a particle is deemed to have collided, a second Monte Carlo step
is needed to select a process.  Since a particle is only permitted to
collide once per time step, this procedure introduces another
constraint, namely
\begin{equation}
\nu_{e,i} \Delta t \ll 1.
\end{equation}
During a collision the momentum and energy of the particle
are appropriately adjusted, and any necessary new particles are
added to the simulation.
Further discussion of Monte Carlo procedures for particle-in-cell
simulations can be found elsewhere
\citep{birdsall_particle--cell_1991,vahedi_monte_1995,donko_particle_2011}.

Each of the codes used in this study implements the basic algorithm
outlined above and discussed in detail in the references.  We have
foregone complications that introduce additional numerical parameters
and perhaps variation in implementation, such as subcycling
\cite{adam_electron_1982}.  Within this framework, each of the codes
has been implemented independently and without consultation between
the authors prior to the benchmarking exercise.  Consequently, the
codes differ in many details.  They are implemented in different
computer languages, use different data structures, are designed for
different computer architectures, and doubtless differ in many other
respects owing to the varied practical and philosophical views of the
authors.  The most salient features of the five codes we have employed
are summarized in table~\ref{tab_progs}.  For convenience of later
reference, each code has been assigned a distinguishing letter.
During the course of the benchmarking, certain minor imprecisions were
discovered, and these have been corrected.  We think it unlikely that
these would have been exposed outside the context of the present
study, since the effects on the results were subtle.  But these were,
nevertheless, implementation errors that came to light as consequence
of benchmarking.  We note also that certain inconsistencies that
emerged during the development of the benchmarks were traced to
deficient pseudo-random number generation.  This was a surprise.
The detailed issues involved are not well understood by us, but we
urge caution when choosing a source of random numbers.

Particle-in-cell calculations are evidently stochastic.  Even in
temporal equilibrium, fluctuations will occur around some average
value.  These fluctuations are driven by the stream of pseudo-random
numbers consumed by the Monte Carlo elements in the simulation, but
may also be affected by other factors, such as round-off error in
finite precision arithmetic.  Pseudo-random number generators
typically are deterministic algorithms initialized with a seed value,
yielding a different sequence of values for every unique seed.
Consequently, every simulation program gives results that depend at
least on the seed value.  Moreover, even computers that nominally
implement standard floating point arithmetic do not always respect the
rounding rules strictly, with the result that the same program
executed on different computers or using different software tools is
likely to give different results.  Consequently, for these and
other reasons, we do not expect that
the diverse implementations available to this study can give identical
results, even for the same physical conditions and numerical
parameters.  We can however ask whether the results from the different
codes can be statistically distinguished.  We approach this problem by
treating the ion densities computed at the mesh points as a set of
random variables.  If we can characterize the density at mesh point
$j$ by an average value $\bar{n}_i(x_j)$ and a standard deviation
$\sigma_i(x_j)$, then, for a particular set of mesh densities
$n_i(x_j)$ drawn from the simulation, we can compute
\begin{equation}
X^2 = \sum_j \frac{\left[n_i(x_j) - \bar{n}_i(x_j)\right]^2}{\sigma_i(x_j)^2}.
    \label{eq_chi2}
\end{equation}
If the random variables are uncorrelated and normally distributed, and
the set of values $n_i(x_j)$ is actually drawn from the distribution
defined by $\bar{n}_i(x_j)$ and $\sigma_i(x_j)$, then the value of
$X^2$ is drawn from a well-known distribution function
\citep[sec. 26.4]{abramowitz_handbook_1965}. If we find that the value
of $X^2$ that we calculate is far into the tail of this probability
distribution, we will conclude that the values $n_i(x_j)$ were likely
not drawn from the assumed distribution.  By this means, we can
establish whether the results of two simulation programs are
statistically distinguishable or not.

\begin{figure}
\includegraphics[width=\columnwidth]{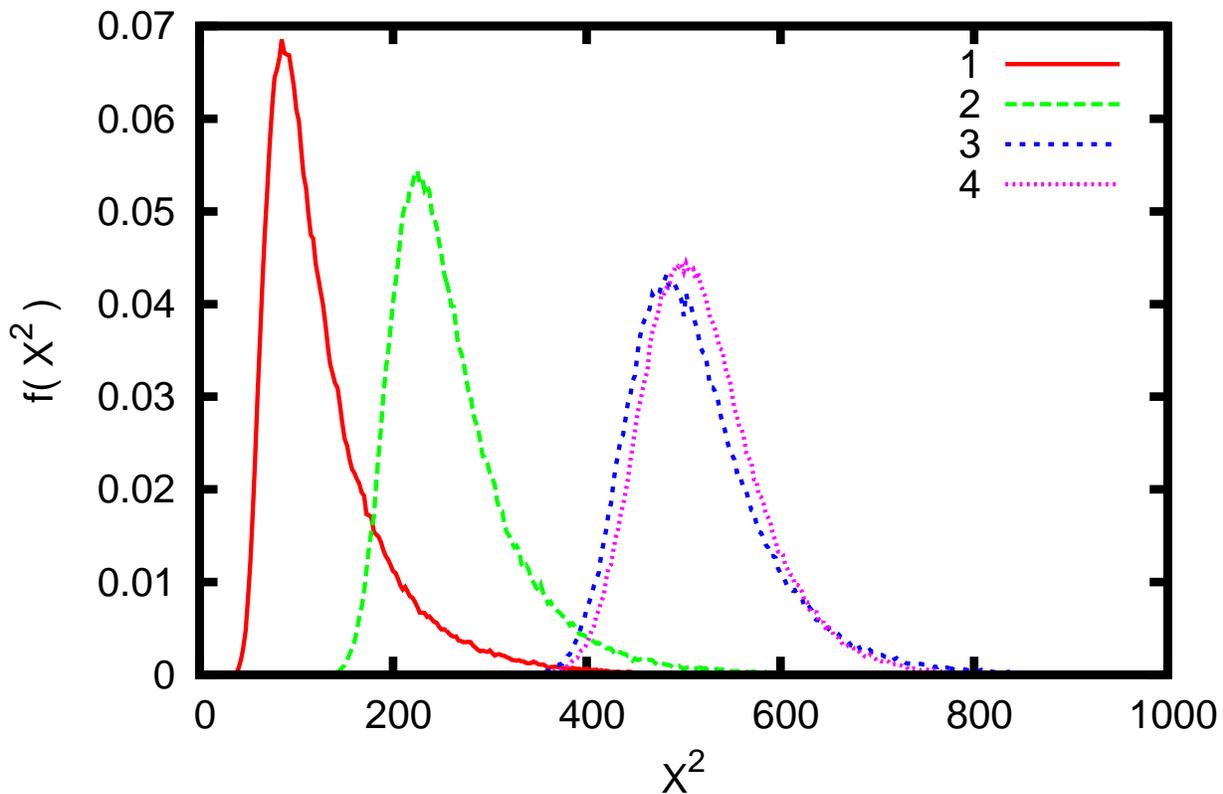}
\caption{Distribution of $X^2$ values for the four benchmark cases
\label{fig_chi_all}}
\end{figure}

There is a difficulty, in that we have no grounds for assuming that the
densities at the mesh points have the properties assumed above of
being uncorrelated and normally distributed.  This, however, only means that
the distribution of values of $X^2$ is not the one usually assumed.  We
can proceed by using one simulation to generate the values
of $\bar{n}_i(x_j)$ and $\sigma_i(x_j)$, and the distribution of $X^2$
values, and we can proceed to test the other simulations against these
results in the manner described above.  We expect a different distribution
of $X^2$ for each benchmark case.  Fig.~\ref{fig_chi_all} shows the
distribution functions that are determined in this way.  Each has been
normalized such that
\begin{equation}
\int_0^\infty f(X^2) dX^2 = 1.
\end{equation}
We note that the conventional $X^2$ distribution is characterized by
the number of degree of freedom, $k$, which is in this case equal to
the number of mesh points.  When $k\gg 1$, the distribution of $X^2$
is approximately normal, with mean value $k$ and standard deviation
$\sqrt{2k}$.  The distributions shown in fig.~\ref{fig_chi_all} indeed
have a mean close to $k$, but evidently have a larger standard
deviation and considerable skew.  We have not investigated the origin
of these features, but we speculate that the cause is correlations
between the density fluctuations at neighbouring mesh points, produced
by plasma dynamical effects.

Our procedure, therefore, begins by generating values of $\bar{n}_i(x_j)$
and $\sigma_i(x_j)$, such that the standard deviation of the mean
values $\bar{n}_i(x_j)$ is negligible compared to the population
standard deviation $\sigma_i(x_j)$.  We obtain these results by
observing fluctuations around a stationary state in an extended calculation
using code E.  From this extended calculation we also find
the $X^2$ distributions shown in fig.~\ref{fig_chi_all}.  We can then
take a result from any of the other codes, compute $X^2$ using
eq.~\ref{eq_chi2}, and refer to the data in fig.~\ref{fig_chi_all} and
table~\ref{tab_chi_comp} to determine the significance of the result.  If
we find no unlikely values, we can declare that the test codes indeed
are statistically indistinguishable.  On the sensitivity of this test,
we can say that a consistent difference between two simulations results of
about 0.5~\% will produce a $X^2$ value that should occur by chance
only about once in 10~000 trials.  A difference of this magnitude
should therefore be easily detected.  On the other hand, a systematic
difference of 0.1~\% shifts the value of $X^2$ by an amount small
compared to the normal ranges indicated in fig.~\ref{fig_chi_all} and
table~\ref{tab_chi_comp}, so that a difference of this magnitude
cannot be discerned.  

If we were concerned to demonstrate only that the test codes give
practically identical results, then we could stop here.  However, we
intend our results to be of value to authors of other kinds of codes
than particle-in-cell codes, and for this reason we need to go a step
further and estimate the numerical errors remaining in our
calculations.  We have done this using a refinement strategy.  Since
the convergence of particle-in-cell simulations with Monte Carlo
collisions, as a function of the numerical parameters, has never been
fully investigated (some indications are give by
\citet{vahedi_capacitive_1993} and \citet{turner_kinetic_2006}) , the
optimal refinement procedure is not clear.  Our procedure is a simple
one.  At each refinement, we halve the time step and the cell size.
Since the algorithm is second order in these quantities, this should
reduce the associated numerical errors by a factor of four.  Numerical
errors associated with the number of particles per Debye length,
$N_D$, vary as $N_D^{-1}$ or $N_D^{-2}$, depending on the
collisionality and nature of the error \cite{turner_kinetic_2006}.  To
achieve a reduction of these errors by at least a factor four, then,
we should increase the number of particles by a factor of four.  This
procedure appears reasonable, but is not guaranteed to reduce the
error in the solution by a factor four, because the relationship
between the velocity space diffusion effects regulated by $N_D$ and
the global error in the solution is not clear.  Nevertheless, by
comparing solutions at different levels of refinement we can form some
view on the magnitude of the errors.

\section{Results}

\begin{figure}
\includegraphics[width=\columnwidth]{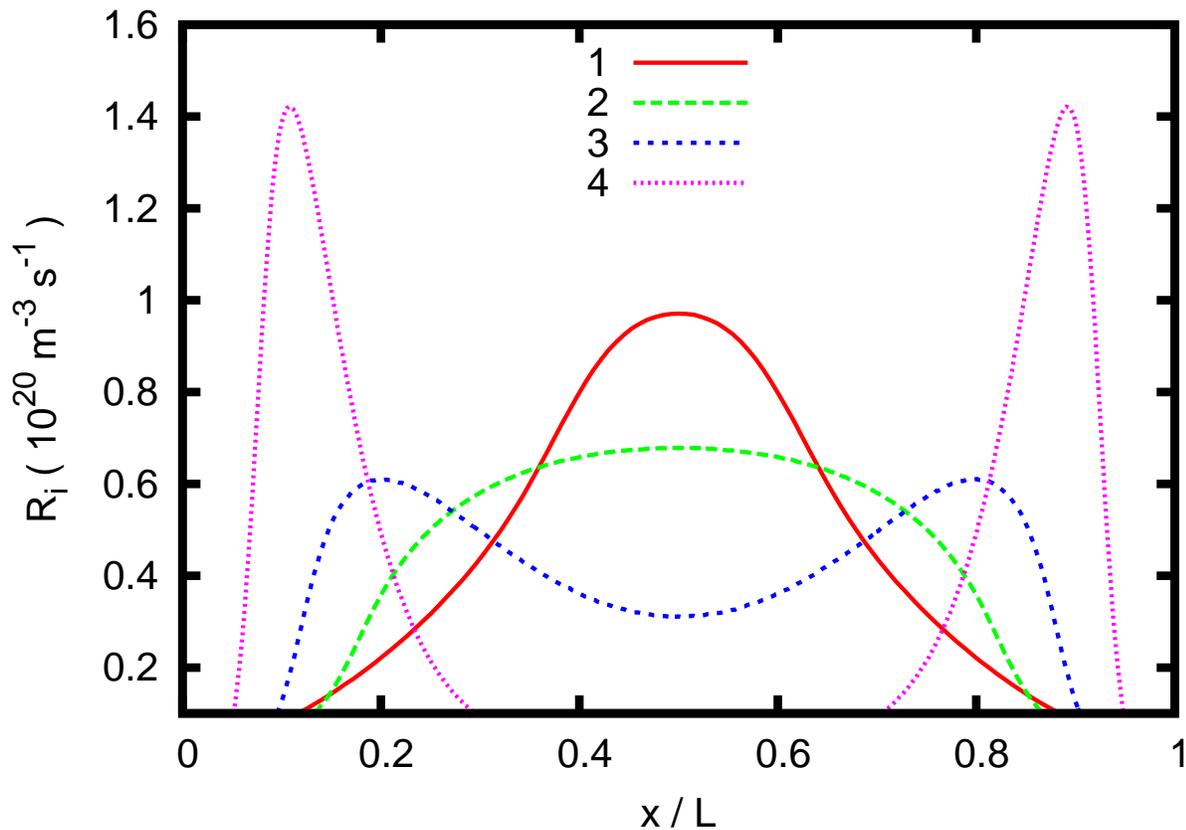}
\caption{Time averaged ionization source term for the four benchmark cases.
\label{fig_ion}}
\end{figure}

\begin{figure}
\includegraphics[width=\columnwidth]{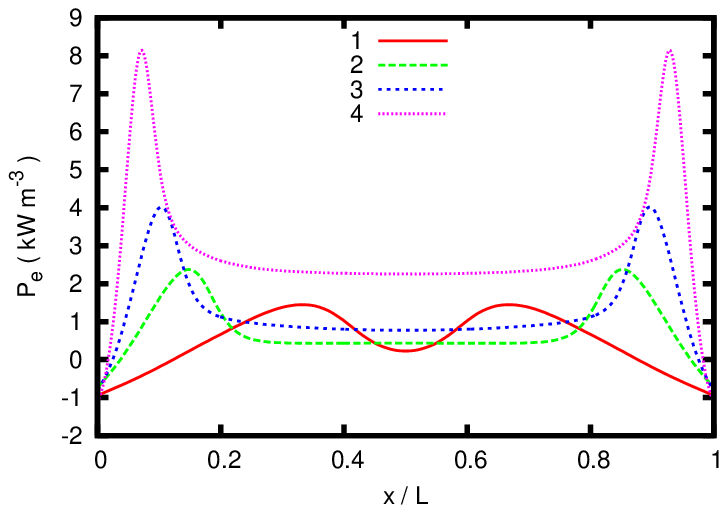}
\caption{Time averaged power density ($\langle  J_e\cdot E\rangle$) coupled to
electrons for the four benchmark cases.\label{fig_je_e}}
\end{figure}

\begin{figure}
\includegraphics[width=\columnwidth]{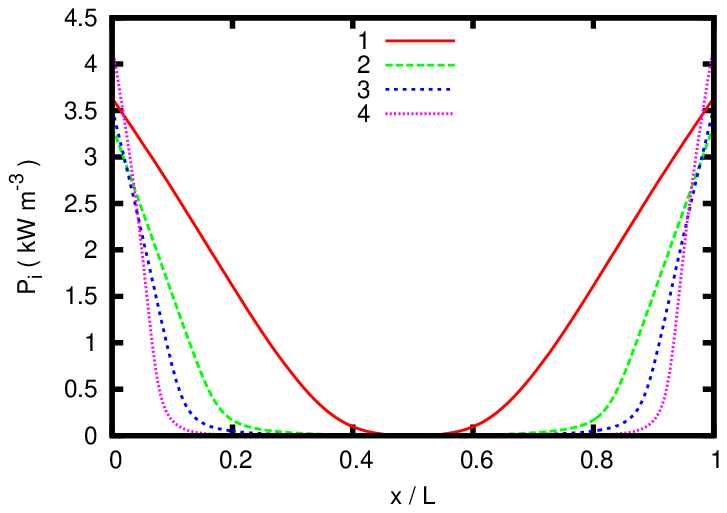}
\caption{Time averaged power density ($\langle  J_i\cdot E\rangle$) coupled to
ions for the four benchmark cases.\label{fig_je_i}}
\end{figure}

\begin{figure}
\includegraphics[width=\columnwidth]{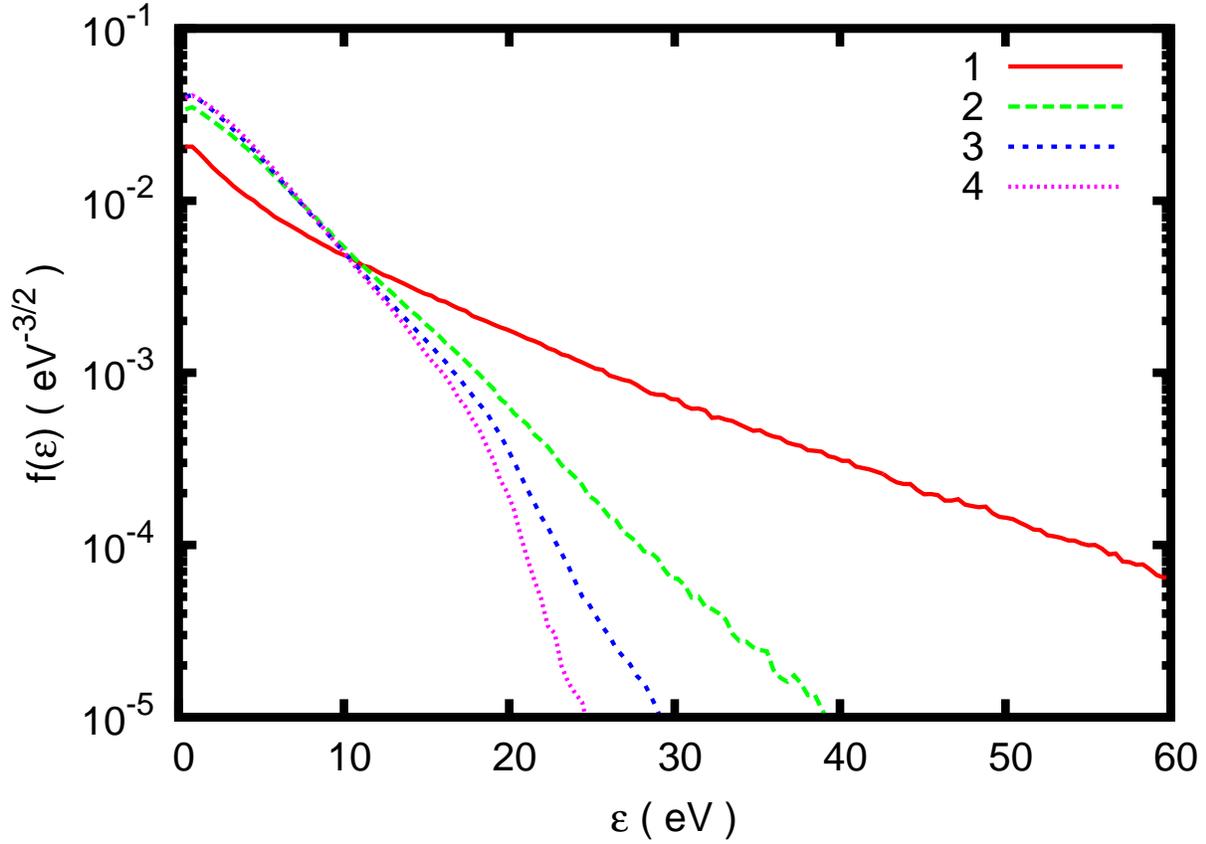}
\caption{Time and space averaged electron energy probability functions
for the four benchmark cases.  These data are normalized such that
$
\int_0^\infty \sqrt{\epsilon} f(\epsilon) d\epsilon = 1.
$
A Maxwell-Boltzmann distribution would appear as a straight line on
this plot.
\label{fig_eedf}}
\end{figure}

A summary of the main physical parameters for each benchmark case,
together with numerical figures of merit, is presented in
table~\ref{tab_char}.  These data show that the benchmark cases
conservatively satisfy the conventional accuracy conditions discussed
above.  The first four figures show the ionization source term
(figure~\ref{fig_ion}), the electrical power coupled to electrons
(figure~\ref{fig_je_e}), the electrical power coupled to ions
(figure~\ref{fig_je_i}) and the electron energy distribution functions
for each of the cases (figure~\ref{fig_eedf}).  We see that as the
pressure increases from case 1 to case 4, both ionization and power
transfer become increasingly concentrated in the sheath region.  Much
of this behaviour is determined by the electron energy relaxation
length.  For electrons below the threshold for inelastic processes,
the energy relaxation length is large compared with the electrode
spacing in all four cases.  Consequently, we expect that the gross
electron mean energy varies rather little in space.  Above the
inelastic threshold, however, the energy relaxation length is larger
than the electrode separation in case 1, but smaller in case 4.  Hence
we find that in case 4, power absorption and dissipation have become
more locally balanced, with maxima in the sheath regions.  The
electron energy distribution functions shown in figure~\ref{fig_eedf}
exhibit no striking structures.  At the lowest pressure, the
distribution function is close to Maxwellian, but at higher pressure
there is an increasing depression of the high energy tail caused by
inelastic collisions.  None of these distribution functions shows in
a marked form any of the
more exotic structures that are sometimes seen, such as bumps, holes
or super-thermal high energy tails \citep{godyak_abnormally_1990,godyak_evolution_1992,godyak_measurement_1992,turner_anomalous_1992}.
These features are commonly
symptomatic of strongly non-local interactions between the electrons
and the fields.  The absence of such interactions means that
simplified models using approximate treatments of the electron
kinetics have a reasonable chance of reproducing the present
benchmarks with tolerable accuracy.

\begin{figure}
\includegraphics[width=0.8\columnwidth]{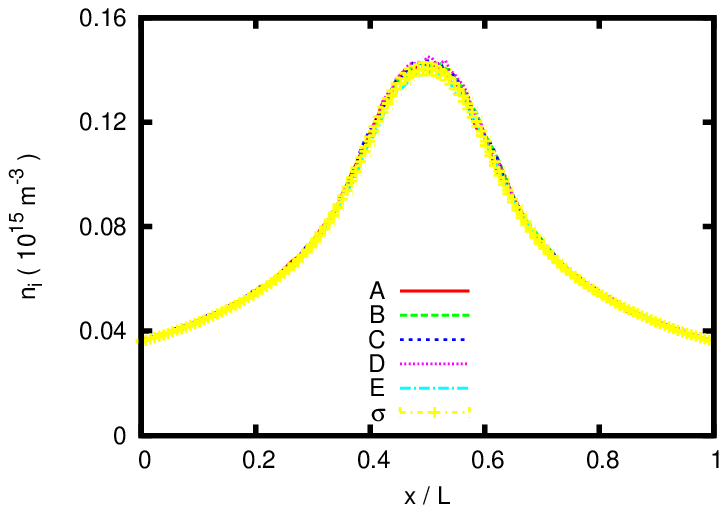}
\includegraphics[width=0.8\columnwidth]{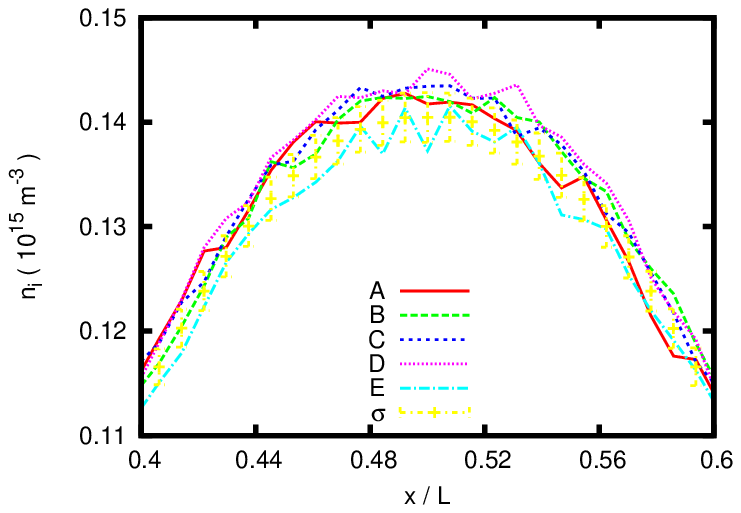}
\caption{Ion density distribution for case 1.  The curves labelled
A to E show the results obtained by the indicated code.  The points
with error bars show the standard deviation obtained from an extended
calculation using code E.
\label{fig_n1}}
\end{figure}

\begin{figure}
\includegraphics[width=0.8\columnwidth]{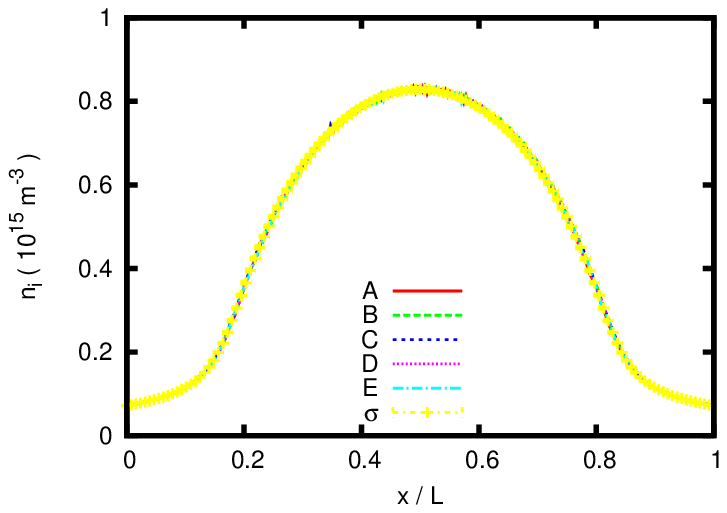}
\includegraphics[width=0.8\columnwidth]{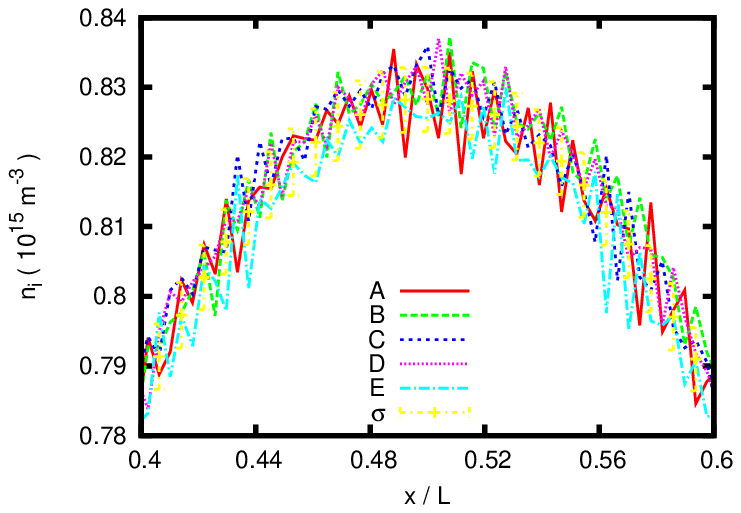}
\caption{Ion density distributions for case 2.  Refer to the caption
of fig.~\ref{fig_n1} for further explanation.
\label{fig_n2}}
\end{figure}

\begin{figure}
\includegraphics[width=0.8\columnwidth]{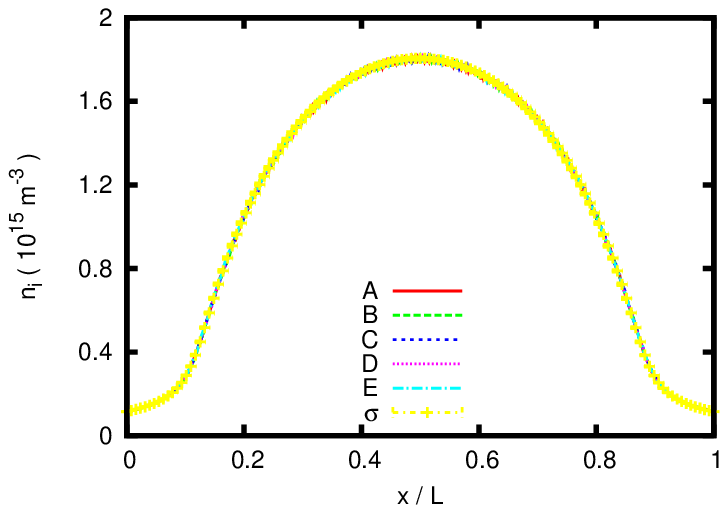}
\includegraphics[width=0.8\columnwidth]{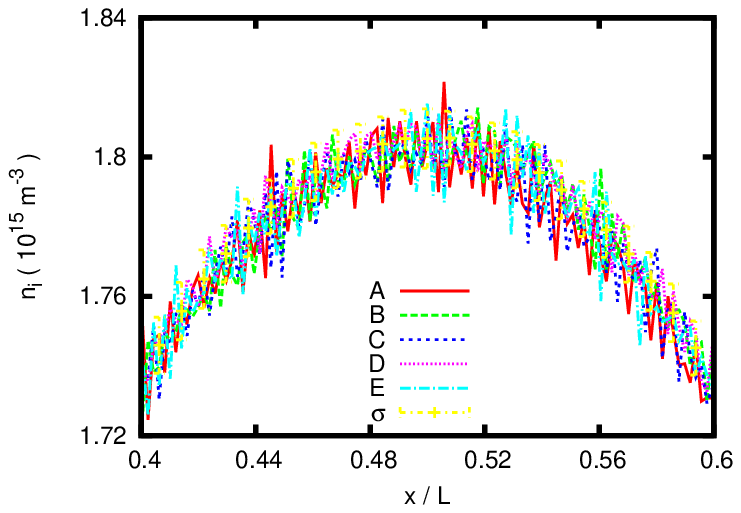}
\caption{Ion densities distributions for case 3. Refer to the caption
of fig.~\ref{fig_n1} for further explanation.
\label{fig_n3}}
\end{figure}

\begin{figure}
\includegraphics[width=0.8\columnwidth]{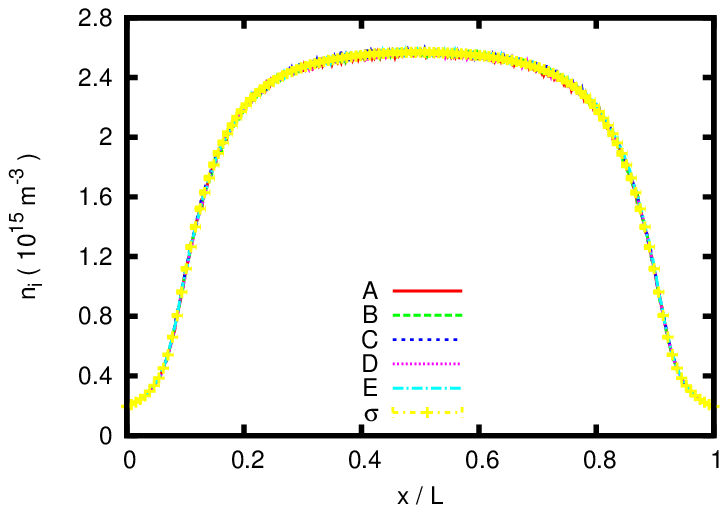}
\includegraphics[width=0.8\columnwidth]{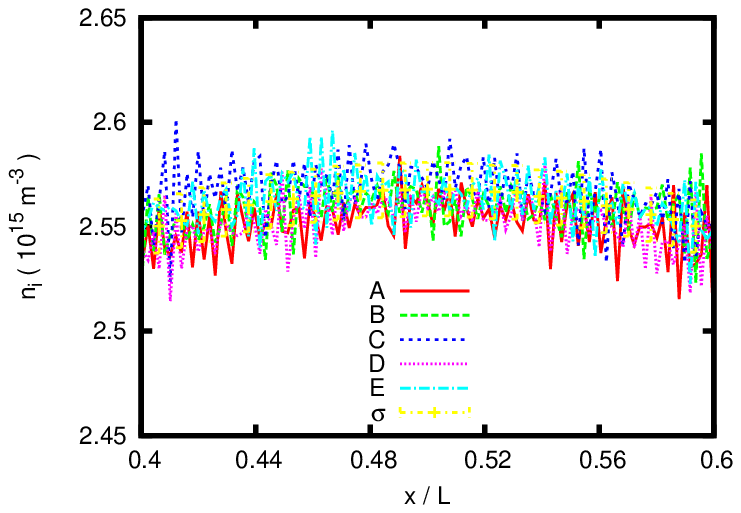}
\caption{Ion density distributions for case 4. Refer to the caption
of fig.~\ref{fig_n1} for further explanation.
\label{fig_n4}}
\end{figure}

In the next group of figures we present the comparisons between the
ion densities for the four benchmark cases.  Figures~\ref{fig_n1},
\ref{fig_n2}, \ref{fig_n3} and \ref{fig_n4} compare the densities
calculated using each of the five codes.  Of course, the densities in
each code fluctuate around some mean value, and we also show in these
figures error bars which denote one standard deviation of this
fluctuation.  All the figures show a global view of the data, and an
expanded region around the discharge mid-plane.  From the global view,
the agreement between the codes is evidently excellent, in the sense
that no disagreement within the error bars can be discerned.  This is
also true of the expanded view, where the size of the fluctuations is
more clearly evident.  As we have suggested above, the coincidence 
of the codes can be examined objectively using a
statistical test.  The results of these statistical tests are
summarized in table~\ref{tab_chi_comp}.  We note that since code E was
used to generated the $X^2$ distributions, the results for that code
are presented essentially as a methodological test.  These data can be
examined in conjunction with the distributions shown in
figure~\ref{fig_chi_all}, but for convenience we have shown in the
table the range of values that can be accepted as being likely to have
occurred by chance. One of the twenty values shown here is outside the
95~\% confidence limits, which is of course the expected outcome.

\begin{figure}
\includegraphics[width=\columnwidth]{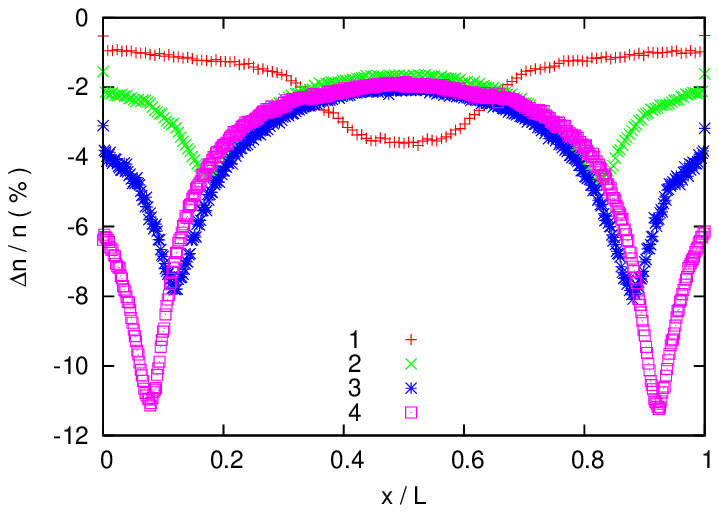}
\caption{Estimation of the residual numerical errors in each of the
four benchmark cases.  These data have been obtained by comparing the
base cases with additional calculations using substantially more
refined numerical parameters.  Since we have compared the solutions at
coincident mesh points, this presentation overstates the significance of
the errors where there are strong spatial gradients ({\em c.f}, figures
\ref{fig_n1},\ref{fig_n2},\ref{fig_n3} and \ref{fig_n4}).\label{fig_err_all}}
\end{figure}

The evidence we have shown above strongly suggests that the codes
under discussion are statistically indistinguishable, and
from this result we wish to urge the conclusion that the codes are
also correct in some usefully strong sense.  Of course, the basis for
this claim is the assumption that five independently developed codes
are unlikely to be united in error.  Even if this is so, the benchmark
results obtained from these codes have an accuracy that is limited by
numerical effects arising from finite time steps, cell sizes and
particle density.  Figure~\ref{fig_err_all} shows evidence relevant to
the question of estimating the size of this residual error.  As we
explained above, we have approached this issue using a sequence of
computations with increasingly refined numerical parameters.  In each
case, the most refined solution has the cell size and time step
reduced by a factor four, and the number of particles increased by a
factor sixteen, for an increase in numerical exertion by a factor of
about sixty four.  By comparing intermediate stages in the refinement
process, we estimate that the numerical uncertainty is reduced by a
factor of approximately ten in the most refined solutions, relative to
the base cases detailed in table~\ref{tab_param}.  In
figure~\ref{fig_err_all} we show the difference between the base case
solutions and the most refined solutions.  As we note in the figure
caption, we have compared the solutions at coincident spatial points,
and this procedure leads to apparently large errors in regions where
there are strong spatial gradients.  A more significant estimator of
the error in each solution is the difference in the mid-plane, which
is seen to be rather uniformly a few percent. The error in the refined
solutions is approximately ten times smaller than this.  These data
provide some guidance to authors of simulation codes other than
particle-in-cell codes as to whether their results are consistent with
the benchmark or not

\section{Discussion}

The results obtained in the present work are appreciably more consistent
than those found in the earlier benchmark comparison of 
\citet{surendra_radiofrequency_1995}.  For example, results from the
three particle-in-cell implementations considered in that exercise
were typically different by $\sim$5~\%, and differences in excess
of 15~\% occurred.  Larger divergences were found between the particle-in-cell
simulations and other kinetic solvers---as much as 100~\% in some cases.
The range of all the simulations considered spanned approximately a factor
of two in electron temperature and a factor of three in density.  On the
evidence available, these differences are not easy to understand---but
the kinetic simulations, at least, are all supposedly solving the same
physical model, so we must assume either implementation error or operator
error ({\em i.e.} inappropriate physical or numerical parameters).
As evidence from other fields shows, errors of these types occur commonly,
and are difficult to eliminate.  In the present work, we cannot claim
with certainty that we have eradicated all such mistakes, but we can
make a limited claim based on the statistical arguments presented above:
Any remaining errors affect the results systematically by appreciably
less than 1~\%.  The residual numerical errors in the base cases
are significantly greater than this, so we have also provided data
using more refined numerical parameters that reduce the overall
precision to about the 1~\% level.  These data provide a
basis for authors of other simulation codes to evaluate the accuracy
of their own results, which of course is one of our central aims.

A precision much better than 1~\% is perhaps of limited practical
value.  In the end, the primary purpose of simulations is to predict
the outcome of experiments, and not many relevant experiments (if any) approach
an accuracy of 1~\%.  Moreover, the accuracy of the simulations is
limited by the available physical data as well as by numerical
considerations.  The most accurate cross section data
available at present probably have an uncertainty of at least a few
percent.  The base case numerical parameters shown in
table~\ref{tab_param} therefore probably entail an appropriate level
of numerical exertion for most purposes.  Indeed we note that the
difference between the present benchmark results and comparable
experiments \citep{godyak_evolution_1992,godyak_measurement_1992} is
far larger than any reasonable estimate of the error due to numerical
effects or faulty cross section data---roughly a factor of two in
density and voltage, for a given current density.  We assume that this
is due to an incomplete physical model.  For example, we have
neglected emission of particles from electrodes and all effects of
excited states \citep{hitchon_physical_1993}.  This topic will be
further explored in future work.

Although the principal aim of the benchmarks that we present here is
to facilitate verification of codes, we note that a secondary
objective can be achieved, and this is to compare the performance of
different implementations.  The interest of this activity is increased
by knowing that the codes are performing the same calculation (within
some tolerance, at least) so that any difference in execution time is
due to hardware and implementation strategy.  The codes under
investigation take different approaches.  Code C, for example, is
implemented in MATLAB, a high level language that generally favours
ease of programming over efficiency.  Codes A and D are traditional
implementations in the C language, and as such similar in concept to
the well-known {\tt PDP1} \citep{verboncoeur_simultaneous_1993}.
Codes B and E both target features of modern computer architectures
such as multi-threaded execution, symmetric multi-processing, and
vectorisation.  These are, relatively speaking, complex
implementations aiming at high performance.  Developing a basis for
verifying such codes was a major motivation of the present work.  On
the benchmark cases discussed here, codes B and E perform comparably,
and are each about a factor of twenty faster than the serial codes A
and D, which are themselves about twice as fast as the MATLAB code C.
These differences are of considerable practical significance.  For
example, when all the calculations are carried out using reasonably
modern desktop computers, benchmark 4 takes approximately six hours to
execute using codes B and E, but almost three weeks using code C.

\section{Concluding Remarks}

Our aim in this work was to develop benchmarks for low-temperature
plasma physics simulations.  We approached this problem by specifying
four benchmark cases, each with comprehensively defined physical and
numerical conditions.  We have then shown that five independently
developed particle-in-cell simulations produce results for these
benchmarks that are indistinguishable on a statistical basis.
A more precise statement would be that the implementation uncertainty
is less than 0.5~\%.  We proceeded to investigate the numerical
uncertainty in the benchmark results, and we have presented a second
set of benchmark results in which the numerical uncertainty is
reduced to approximately the same level as the implementation
uncertainty.  Thus we claim that these latter results have an uncertainty
at about the 1~\% level.  We do not think that there is any advantage
in attempting to further refine these results at the present time,
allowing for the uncertainty in both basic data and experimental
characterization.  Our intention is that these data can be used
by developers of particle-in-cell codes to verify their work in a
rather rigorous fashion, using the statistical procedures that we have
discussed, while authors of codes based on other principles can
evaluate the accuracy of their work by less formal methods.

Of course, the present benchmarks exercise only a subset of the features
likely to be desired in comprehensive simulation packages.  For example,
we have not treated multi-dimensional effects or electro-magnetic
effects, and plasma chemistry only in a limited fashion.  Nevertheless,
we think that the methodology we have developed represents a powerful
platform for future developments encompassing these more advanced
aspects.

\begin{acknowledgments}

The authors thank G J M Hagelaar and L C Pitchford for insightful
discussions, in particular (but not only) on matters related to
computation of transport coefficients.  MMT and SJK acknowledge the support
of Science Foundation Ireland under grant numbers 07/IN.1/I907 and
08/SRC/I1411. AD and ZD thank P.\ Hartmann for his contributions to
the code development, and the Hungarian Fund for Scientific Research
for the support provided through grants K77653 and K105476. DE and TM
acknowledge the support of the German Research Foundation DFG in the
frame of the Collaborative Research Centre TRR 87.

\end{acknowledgments}

\bibliography{zotero}

\begin{table}
\begin{tabular}{lllllll}
Physical parameters \\
\hline
  & & & 1 & 2 & 3 & 4 \\
Electrode separation & $L$ & ( cm ) & \multicolumn{4}{c}{6.7} \\
Neutral density & $N$ & ( 10$^{20}$ m$^{-3}$ ) & 9.64 & 32.1 & 96.4 & 321 \\
Neutral temperature & $T_n$ & ( K ) & \multicolumn{4}{c}{300} \\
Frequency & $f$ & ( MHz ) & \multicolumn{4}{c}{13.56} \\
Voltage & $V$ & ( V ) & 450 & 200 & 150 & 120 \\
Simulation time & $t_S$ & ( s ) & $1280/f$ & $5120/f$ & $5120/f$ & $15360/f$\\
Averaging time & $t_A$ & ( s ) & $32/f$ & $32/f$ & $32/f$ & $32/f$\\
\\
Physical constants \\
\hline
  & & & 1 & 2 & 3 & 4 \\
Electron mass & $m_e$ & ( 10$^{-31}$ kg ) &  \multicolumn{4}{c}{9.109} \\
Ion mass & $m_i$ & (  10$^{-27}$ kg ) & \multicolumn{4}{c}{6.67} \\
\\
Initial conditions \\
\hline
  & & & 1 & 2 & 3 & 4 \\
Plasma density & $n_0$ & ( 10$^{14}$ m$^{-3}$ ) & 2.56 & 5.12 & 5.12 & 3.84 \\
Electron temperature & $T_e$ & ( K ) &  \multicolumn{4}{c}{30 000} \\
Ion temperature & $T_i$ & ( K ) &  \multicolumn{4}{c}{300} \\
Particles per cell & $N_C$ & & 512 & 256 & 128 & 64 \\
\\ 
Numerical parameters \\
\hline
  & & & 1 & 2 & 3 & 4 \\
Cell size & $\Delta x$ & ( m ) & $L/128$ & $L/256$ & $L/512$ & $L/512$ \\
Time step size & $\Delta t$ & ( s ) & $(400 f)^{-1}$ & $(800 f)^{-1}$ & $(1600 f)^{-1}$ & $(3200 f)^{-1}$ \\
Steps to execute & $N_S$ & & 512 000 & 4 096 000 & 8 192 000 & 49 152 000 \\
Steps to average & $N_A$ & & 12 800 & 2 5600 & 51 200 & 102 400
\end{tabular}
\caption{Physical and numerical parameters for the benchmarks\label{tab_param}}
\end{table}

\begin{table}
\begin{tabular}{llll}
Code & Author(s) & Language & Architecture \\
\hline
A    &  Derzsi and Donk\'o & C           & CPU \\
B    &  Eremin             & C for CUDA  & GPU \\
C    &  Lafleur            & MATLAB      & CPU \\
D    &  Mussenbrock        & C           & CPU \\
E    &  Turner             & C           & CPU (multithreaded) 
\end{tabular}
\caption{Summary characteristics of the five simulation programs
employed in this study.\label{tab_progs}}
\end{table}


\begin{table}
\begin{center}
\begin{tabular}{llllll}
Physical characteristics \\
\hline
                           & 1       & 2        & 3        & 4        \\
$n_i$ ( 10$^{15}$ m$^{-3}$ ) & 0.140   & 0.828    & 1.81     & 2.57      \\
$k_B T_e$ ( eV )           & 9.36    & 4.69     & 3.95     & 3.65      \\ 
$S_e$ ( W m$^{-2}$ )        & 34.3    & 51.6     & 85.2     & 193       \\
$S_i$ ( W m$^{-2}$ )        & 90.6    & 43.3     & 32.0     & 27.1      \\
$J_i$ ( A m$^{-2}$ )        & 0.219   & 0.215    & 0.195    & 0.186     \\
\\
Numerical characteristics  \\
\hline
                           & 1       & 2        & 3        & 4        \\
$\omega_p\Delta t$         & 0.121   & 0.150    & 0.110    & 0.066     \\
$\lambda_D / \Delta x$     & 3.72    & 2.14     & 2.66     & 2.14      \\
$\nu_e \Delta t$           & 0.0158  & 0.0262   & 0.0391   & 0.0643    \\
$\nu_i \Delta t$           & 0.00688 & 0.0114   & 0.0171   & 0.0283    \\
$N_D$                      & 1042    & 886      & 1204     & 917       \\
$N_P$                      & 31900   & 118000   & 283000   & 329000   
\end{tabular}
\end{center}
\caption{Physical and numerical characteristics for the four benchmark
cases.  In the upper section of the table, $n_i$ and $T_e$ are the
ion density and electron temperature in the mid-plane of the
discharge, $S_e$ and $S_i$ are the line integrated electrical power
coupled to electrons and ions, respectively, and $J_i$ is the
ion current collected at either electrode.  All these quantities are
time averaged.  The lower table shows numerical figures of merit,
which are evaluated at the mid-plane of the discharge using time averaged
data, apart from the total number of particles $N_P$, which figure refers
to the entire discharge volume.\label{tab_char}}
\end{table}

\begin{table}
\begin{center}
\begin{tabular}{lllll}
      & 1       & 2        & 3        & 4        \\
\hline
A     & 240     & 199      & 540      & 606      \\
B     & 192     & 310      & 503      & 543      \\
C     & 328     & 408      & 503      & 703      \\
D     & 242     & 209      & 425      & 517      \\
E     & 57      & 219      & 592      & 542      \\
\hline
95\%  & 55--303 & 177--435 & 405--693 & 417--665 \\
99\%  & 48--405 & 160--548 & 382--798 & 392--730 
\end{tabular}
\end{center}
\caption{$X^2$ values for each of the simulation programs applied to
each of the benchmark cases.  The lines at the bottom of the table
show the ranges of values that encompass 95~\% and 99~\% of the
area under the distributions shown in figure~\ref{fig_chi_all}.
\label{tab_chi_comp}}
\end{table}

\appendix

\section{Moment models}
\label{ap_moment}
In this appendix we simulate the benchmark conditions using a moment
model, and we compare the data so obtained with the particle-in-cell
simulation results presented above.  A moment model is based a greatly
simplified physical model.  Instead of calculating the phase space
distribution functions of the particles, as a particle-in-cell
simulation does, a moment model solves conservation equations for a
limited number of macroscopic physical quantities, such number
density, momentum, and thermal energy.  These equations incorporate a
set of rate constants and transport coefficients, which are in
principle dependent on the energy or speed distributions of the
particles.  Moreover, the moment equations are themselves developed by
computing velocity space moments of the Boltzmann equation, and this
procedure leads in principle to an infinite set of mutually coupled
partial differential equations.  If we solve for only the first two or
three moments (as implied above) then this hierarchy of equations must
be truncated by some means---the so-called ``closure problem.''  So
although a solution of the moment equations is computationally
economical, compared to solving the Boltzmann equation, there is
considerable underlying theoretical complexity.  Indeed most moment
models include informally developed elements, such as transport and
rate coefficients injected from a Monte Carlo simulation, a Boltzmann
equation solution, or experimental data.  There may also be
differences in detail on the formulation of the moment equations
themselves, the implementation of boundary conditions and the choice
of numerical procedure. How the many decisions
involved in designing a moment model affect the accuracy
of the results is not well understood, and this is an issue that
benchmarking can address.  Our intention here, however, is more
modest.  We aim only to offer an example of applying a moment model to
the benchmark cases described in the main text.  The moment model that
we have employed is a commercial one, offered by COMSOL Multiphysics
\citep{comsol_multiphysics_2012}.  In most respects, this is a
conventional formulation, calculating three moments for electrons and
two for ions.  An unusual feature is that the species densities are
expressed in a logarithmic form---details of this aspect and others
are to be found in documentation for the package, and will not be
discussed here.  For electrons, rate constants and transport
coefficients were computed from the cross sections specified above
using BOLSIG+ \citep{hagelaar_solving_2005}, while the ion mobility
$\mu_i$ was expressed using the result given by
\citet{patterson_temperature_1970} for a gas temperature of 300~K:
\begin{equation} 
\mu_i N = 2.69\left[1 + 1.2\times 10^{-3}(E/N)^2 +
4.2\times 10^{-8} (E/N)^4\right]^{-\frac{1}{8}}
\end{equation}
where $E/N$ is expressed in Td.  This result is in reasonable
agreement (within 5~\% over the range 1~Td $< E/N < 900$~Td) with
Monte Carlo transport calculations using the ion scattering cross
sections given above.  Naturally, the numerical parameters specified
for the particle-in-cell simulations are inapplicable to the moment
model, which has different stability and accuracy conditions.

\begin{figure}
\includegraphics[width=0.8\columnwidth]{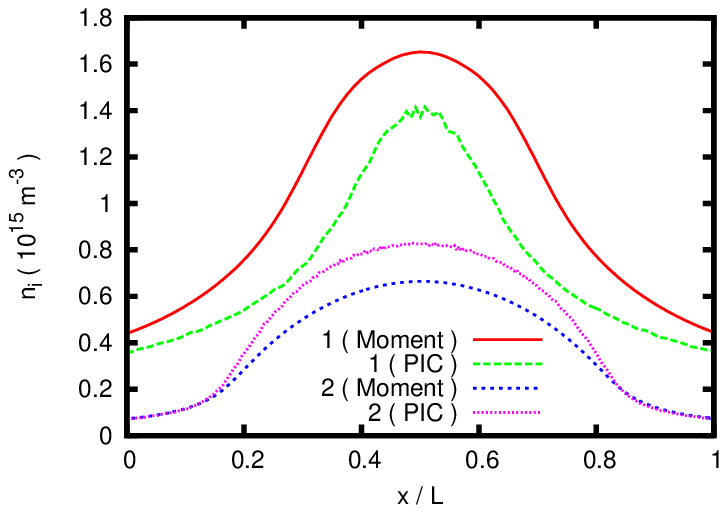}
\includegraphics[width=0.8\columnwidth]{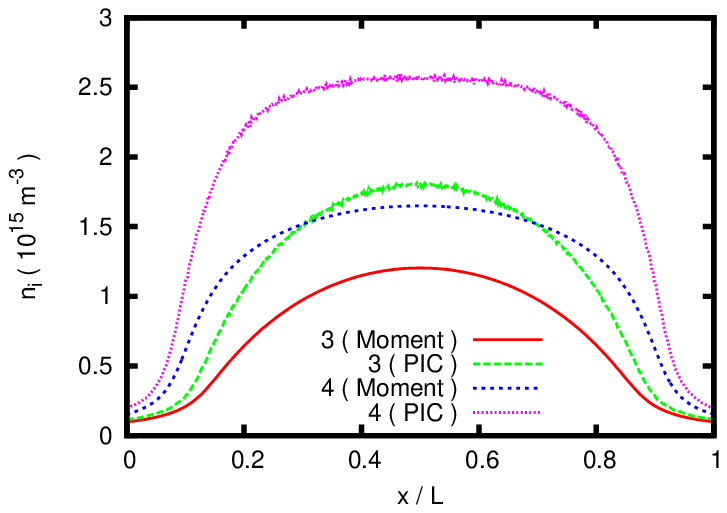}
\caption{Comparison of the moment model computations described in
\ref{ap_moment} with the particle-in-cell calculations
of the main text.  The top panel shows benchmark cases 1 and 2, with the data
for case 1 scaled by a factor of ten for clarity of presentation, and the
bottom panel shows benchmark cases 3 and 4.
\label{fig_moment}}
\end{figure}

Comparisons between these moment calculations and the particle-in-cell
simulation data are shown in figure~\ref{fig_moment}. In general,
there is qualitative agreement concerning the spatial distribution of
the ion density, but disagreement about the maximum density.  Perhaps
surprisingly, this disagreement increases with neutral gas pressure,
and is about 50~\% in case 4.  Such variations cannot reasonably be
attributed to numerical effects, and we assume that modelling issues
are in play, such as the procedure for computing rate and transport
coefficients, use of the drift-diffusion approximation, {\em etc.}
The trend of these differences is generally similar to 
observations in the earlier benchmark of
\citet{surendra_radiofrequency_1995}, where the moment models
generally gave smaller densities than the particle-in-cell
simulations.  A more comprehensive enquiry would be of interest.  This
is beyond the scope of the present study, but may be the subject of
future work.

\end{document}